\documentclass[journal]{IEEEtran}

\IEEEoverridecommandlockouts

\usepackage{graphics} 
\usepackage{epsfig} 
\usepackage{mathptmx} 
\usepackage{times} 
\usepackage{amsmath} 
\usepackage{amssymb}  
\usepackage{multirow}
\usepackage[ruled]{algorithm2e}
\usepackage{epstopdf} 
\usepackage{subfigure} 
\usepackage{stfloats} 
\usepackage{bm}
\usepackage{color}
\usepackage{amsthm}
\usepackage{cite}
\usepackage{array}
\usepackage{threeparttable}
\usepackage{nomencl}
\usepackage{braket}
\usepackage[switch,pagewise,columnwise]{lineno}

\usepackage{booktabs} 
\usepackage{multirow} 
\usepackage{makecell} 
\usepackage{threeparttable}

\makenomenclature

\IEEEoverridecommandlockouts    

\newcounter{algsubstate}

\title{Tomography of Quantum States from Structured Measurements via quantum-aware transformer}

\author{Hailan~Ma, Zhenhong~Sun, Daoyi~Dong, Chunlin~Chen, Herschel Rabitz
	\thanks{This work was supported by the Australian Research Council’s Future
		Fellowship funding scheme under Project FT220100656 and the National Natural Science Foundation of China No.62073160. H. Rabitz acknowledged the ARO (W911NF-19-1-0382) for algorithmic aspects of the research and DOE (DE-FG02-02ER15344) for research on testing of the procedure (Corresponding authors: Zhenhong Sun, Daoyi Dong).}%
	\thanks{H. Ma is with the School of Engineering 
		and Technology, University of New South Wales, Canberra, ACT 2600, Australia, and is also with  the School of  Engineering, The Australian National University, Canberra, ACT 2601, Australia (email: hailanma0413@gmail.com).}
	\thanks{Z. Sun is with the School of  Engineering, The Australian National University, Canberra, ACT 2601, Australia (email: zhenhongsun1992@outlook.com).}
	\thanks{D. Dong is with the Australian Artificial Intelligence Institute, Faculty of
		Engineering and Information Technology, University of Technology Sydney,
		Sydney, NSW 2007, Australia (email: daoyidong@gmail.com)}
	\thanks{C. Chen is with the Department of Control and Systems Engineering, School of Management and Engineering, Nanjing University, Nanjing 210093, China (e-mail: clchen@nju.edu.cn).}
	\thanks{H.  Rabitz is with the  Department of  Chemistry,  Princeton University, Princeton, NJ 08544, USA (e-mail: hrabitz@princeton.edu).}
}

\begin{document}
	\maketitle

	
	\begin{abstract}
		
		Quantum state tomography (QST) is the process of reconstructing the state of a quantum system (mathematically described as a density matrix) through a series of different measurements, which can be solved by learning a parameterized function to translate experimentally measured statistics into physical density matrices. However, the specific structure of quantum measurements for characterizing a quantum state has been neglected in previous work. In this paper, we explore the similarity between highly structured sentences in natural language and intrinsically structured measurements in QST. To fully leverage the intrinsic quantum characteristics involved in QST, we design a quantum-aware transformer  (QAT) model to capture the complex relationship between measured frequencies and density matrices. In particular, we query quantum operators in the architecture to facilitate informative representations of quantum data and integrate the Bures distance into the loss function to evaluate quantum state fidelity, thereby enabling the reconstruction of quantum states from measured data with high fidelity. Extensive simulations and experiments (on IBM quantum computers) demonstrate the superiority of the QAT in reconstructing quantum states with favorable robustness against experimental noise.

	\end{abstract}


	\section{Introduction}\label{sec:introduction}
	
\begin{figure}[t]
		\centering	
		\includegraphics[width=0.95\linewidth]{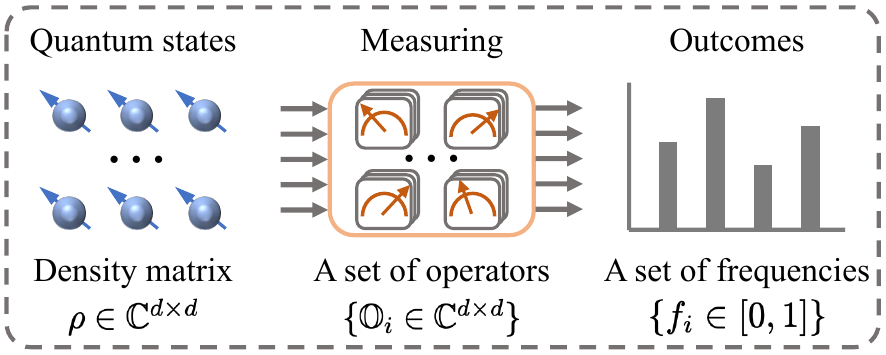}
		\caption{Physical process of QST. Given a number of identical copies of a quantum state (denoted as $\rho \in \mathbb{C}^{d\times d}$, with $d$ representing system dimension) (see Section~\ref{sec:preQST} for detailed information), a set of measurements (with each measurement operator denoted as $\mathbb{O}_i\in \mathbb{C}^{d\times d}$) are performed on quantum states, with the collected outcomes represented into a set of frequencies $\{f_i\in[0,1]\}$.}
		\label{fig:nnqst}
	\end{figure}

	Modern quantum technologies exploit distinctive features of quantum systems to achieve performance beyond classical models. This advantage relies on the capability to manipulate~\cite{dong2010quantum,dong2022quantum,dong2023learning}, and measure quantum states~\cite{arenz2020drawing,yu2020hybrid}. To verify and benchmark those quantum tasks, quantum state tomography, aiming at reconstructing an unknown state via measurements has been recognized as a fundamental ingredient for quantum information processing~\cite{ghosh2020reconstructing}. To reconstruct a quantum state (mathematically described as a density matrix, i.e., a positive-definite Hermitian matrix with unit trace), one may first perform measurements on a collection of identically prepared copies of a quantum system, gathering statistical outcomes to a set of frequencies (see Fig. \ref{fig:nnqst} for detailed information).  However, the data collection only provides a probabilistic result~\cite{qi2013quantum}, and the precise characterization of quantum states hinges on the capability of inferring the density matrix using appropriate estimation algorithms.  
	
	The advancements in ML have led to a great upsurge of research in quantum technology~\cite{dong2019differential,ma2023compression,ma2023curriculum,chen2016quantum,wu2019learning,zhou2025auxiliary}. In particular, neural networks (NNs) have demonstrated their intrinsic capability of efficiently representing quantum states~\cite{torlai2020machine,carleo2017solving,neugebauer2020neural} and have been widely investigated in QST tasks~\cite{wang2022ultrafast,lohani2020cnn,schmale2022cnn,pan2022cnn,cha2021attention,ma2024neural,lohani2023dimension,neugebauer2020neural,an2024unified,palmieri2024enhancing,ma2025learning}. Benefitting from different architectures, NNs are now actively explored for learning quantum states. Fully connected neural networks (FCNs) were first adopted to approximate a function mapping measured frequencies to physical density matrices~\cite{xu2018neural} and to realize an ultrafast reconstruction of 11-qubit states under noise~\cite{wang2022ultrafast}. FCNs have demonstrated their capability of denoising the state-preparation-and-measurement errors~\cite{palmieri2020experimental} and their generalization and robustness in dealing with limited measurement resources~\cite{Ma2021CDC,ma2024neural}. 
	Owing to their capability in dealing with local correlations in spatial feature maps, convolutional neural networks (CNNs) have been utilized to reconstruct 2-qubit quantum states in incomplete measurements~\cite{lohani2020machine,danaci2021machine,lohani2021experimental} and have introduced to benchmark quantum tomography completeness and fidelity without explicitly carrying out state tomography~\cite{teo2021benchmarking}. 
	
	\begin{figure*}[htp]
		\centering
		\includegraphics[width=0.9\textwidth]{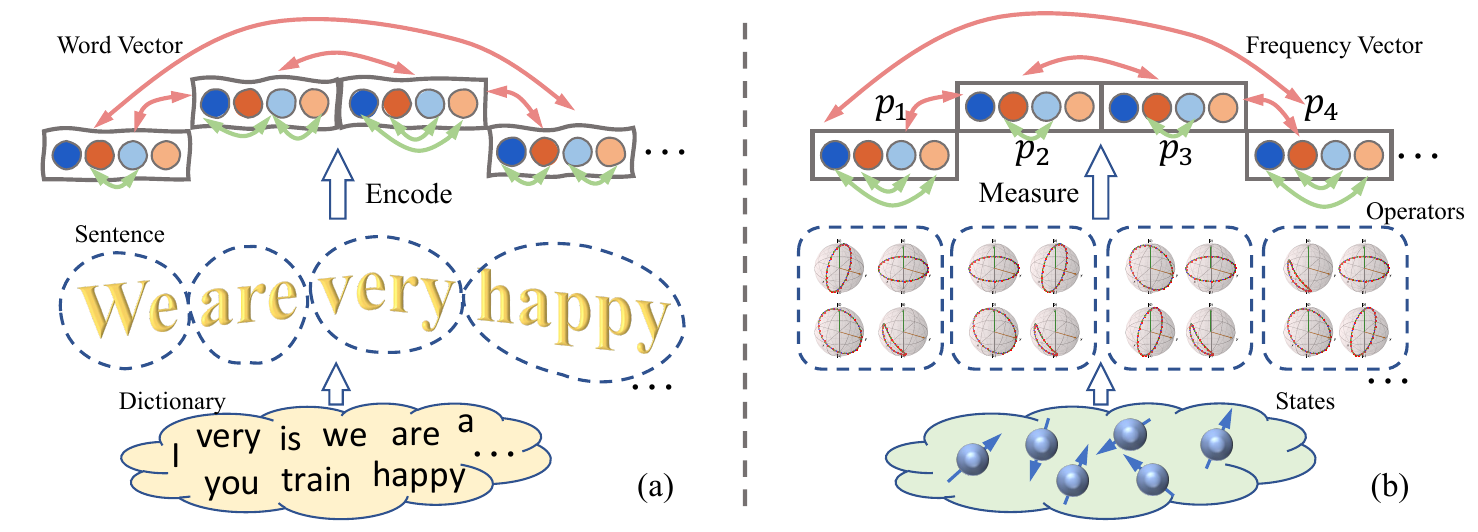}
		\caption{Simalirity between languaging model using words and characters, and QST using structured measurements. (a) A sentence is encoded using four words, with each word containing several characters. The character correlations come in two characters in one word or two characters from different words. Although the basic element is a character, words are encoded into vectors for further assessment. (b) To specify a quantum state, several sets of measurements are performed, with each representing a detector that contains a set of positive semi-definite matrices that sum to identity. The measured proabilities from four detectors are represented into four vectors, which can be further utilized for state reconstructions.}
		\label{fig:lang}
	\end{figure*}
	
	Recently, there have been attempts to explore sequential information among quantum data~\cite{carrasquilla2019reconstructing,zhu2022flexible}. The attention has been utilized to model quantum entanglement across an entire quantum system that enables one to learn probability function~\cite{cha2021attention,zhong2022quantum}.
	However, the specific structure of quantum measurements for characterizing a quantum state (more precisely, its density matrix) has been neglected in previous work. Recall that QST aims at reconstructing physical density matrices from experimentally measured data, where a set of measurements with special patterns (structures) are involved. Such a process can be represented in a structured model that is similar to language modeling where a sentence is composed of words with each one containing several characters (refer to Fig. \ref{fig:lang} for detailed information). To fully leverage the intrinsic quantum characteristics involved in QST, we design a quantum-aware transformer (QAT) model to capture the complex relationship between experimentally measured frequencies and physical density matrices. Different from~\cite{cha2021attention,zhong2022quantum} that utilize attention-based NNs to learn the probability from the direct one-shot measured data, our approach focuses on translating experimentally observed frequencies into physical density matrices directly. In particular, we query quantum operators into the architecture to facilitate informative representations of quantum data and integrate the Bures distance into the loss function to approximate the quantum state fidelity~\cite{marian2008bures}, thereby enabling the reconstruction of quantum states with high fidelity. Extensive simulations in different scenarios verify the superiority of the QAT-QST approach over two machine learning-based QST methods (FCN and CNN) and two traditional QST methods (linear regression estimation and maximum likelihood estimation). Experiments on IBM quantum devices further strengthen the potential of our proposed method in reconstructing quantum states from experimentally measured data. Those results thus highlight the remarkable capability of a quantum-aware transformer in reconstructing quantum states with high fidelity.
	

	The main contributions of our work are summarized as 
	\begin{itemize}
		
		\item We explore the similarity between representing sentences using words composed of characters and representing quantum states using structured quantum measurements (i.e., quantum detectors) composed of several measurement operators (refer to Fig. \ref{fig:lang} for comprehensive details). This analysis offers valuable insight into effectively harnessing the transformer model for QST. 
		\item We design a quantum-aware transformer model that leverages the intrinsic quantum characteristics, including the query of quantum operators for informative representation learning of quantum states and the utilization of Bures distance for precise evaluation of quantum state fidelity. This model is built to translate experimentally measured frequencies into physical density matrices while maintaining a high level of fidelity. 
		
		\item We conduct comprehensive simulations and experiments, utilizing IBM quantum computers, to assess the capabilities of the proposed quantum-aware transformer in reconstructing quantum states. Our findings demonstrate its remarkable robustness against experimental noise, showcasing superior performance compared to both traditional methods and other ML-based approaches.
	\end{itemize}
	
The rest of this paper is organized as follows. Section \ref{Sec:pre} introduces several basic concepts about neural network architectures and QST. In Section \ref{Sec:method}, the proposed QAT-QST method is presented in detail. Numerical and experimental results of different cases are provided in Section \ref{Sec:results}.  Concluding remarks are given in Section \ref{Sec:conclusion}.

\section{Preliminaries}\label{Sec:pre}
This section introduces basic concepts about QST and QST using NNs.

\subsection{Quantum state tomography}\label{sec:preQST}

In the physical world, a quantum state is usually denoted as a unit complex vector $|\psi\rangle$ which can also be viewed as a column vector $v$. Its adjoint is denoted as $\langle\psi|$ ($v^\dagger$), which corresponds to the conjugate and transpose of $v$. The simplest nontrivial Hilbert space is two-dimensional, upon which the quantum system is called a \textbf{qubit}. An orthonormal basis of a qubit system is usually denoted as $|0\rangle$ and $|1\rangle$, corresponding to the classical 0-1 bit and forming the basic unit of quantum information. States that can be represented by a single vector are called \textbf{pure states}. When a quantum system is in a statistical ensemble of different state vectors with proper possibilities, the state of this system would be a \textbf{mixed state} and cannot be described with a single vector. In this case, a mixed state is usually denoted as a \textbf{density matrix} $\rho$
\begin{equation}
	\rho = \sum_i p_i |\psi_i\rangle \langle \psi_i|,
	\label{eq:densitymatrix}
\end{equation}
where $p_i$\label{symbpi} are real coefficients that represent the classical probability of the system is $|\psi_i\rangle$. Generally, a density matrix $\rho$ is a positive-definite Hermitian matrix with a unit trace, representing three numeric attributes: 
(i) $\rho=\rho^{\dagger}$ (where $\rho^{\dagger}$ is the conjugate and transpose of $\rho$), (ii) $\textup{Tr}(\rho)=1$, and (iii) $\rho \geq 0$. QST is a process of deducing the density matrix (denoted as $\rho$) of a quantum state based on a set of measurements (denoted as Hermitian operators $\mathbb{O}$). Measurement operators are usually positive operator-valued measures (POVMs) and can be represented as a set of positive semi-definite matrices ${\mathbb{O}_i}$ that sum to identity ($\sum_i \mathbb{O}_i=\mathbb{I}$)~\cite{wiseman2009quantum}. 

According to Born's rule,  the true probability of obtaining the outcome $i$ when measuring the quantum state $\rho$ with the measurement operator $\mathbb{O}_i$ is calculated as 
\begin{equation}
	p_i = \textup{Tr}(\mathbb{O}_i\rho).
	\label{eq:prob}
\end{equation}
However, in practical implementations, this value is not accessible. As demonstrated in Fig. \ref{fig:nnqst}, one may perform a set of measurements on a number of identical copies of the quantum system, with outcomes gathered as frequencies. Given $N_t$ physical systems for measurements,  statistical data are gathered to obtain the following frequencies 
\begin{equation}
	f_i=\frac{n_i}{N_t},
	\label{eq:frequency}
\end{equation}
with $n_i$ representing the occurrences for the outcome $\mathbb{O}_i$. Here, $f_i$ is a statistical approximation to $p_i$~\cite{qi2013quantum} and the goal of QST is to deduce a density matrix $\rho$ based on the measured statistics $\{f_i\}$.

\subsection{QST using NNs}\label{Subsec:NN-QST} 	
\begin{figure}[ht]
	\centering
	\includegraphics[width=0.45\textwidth]{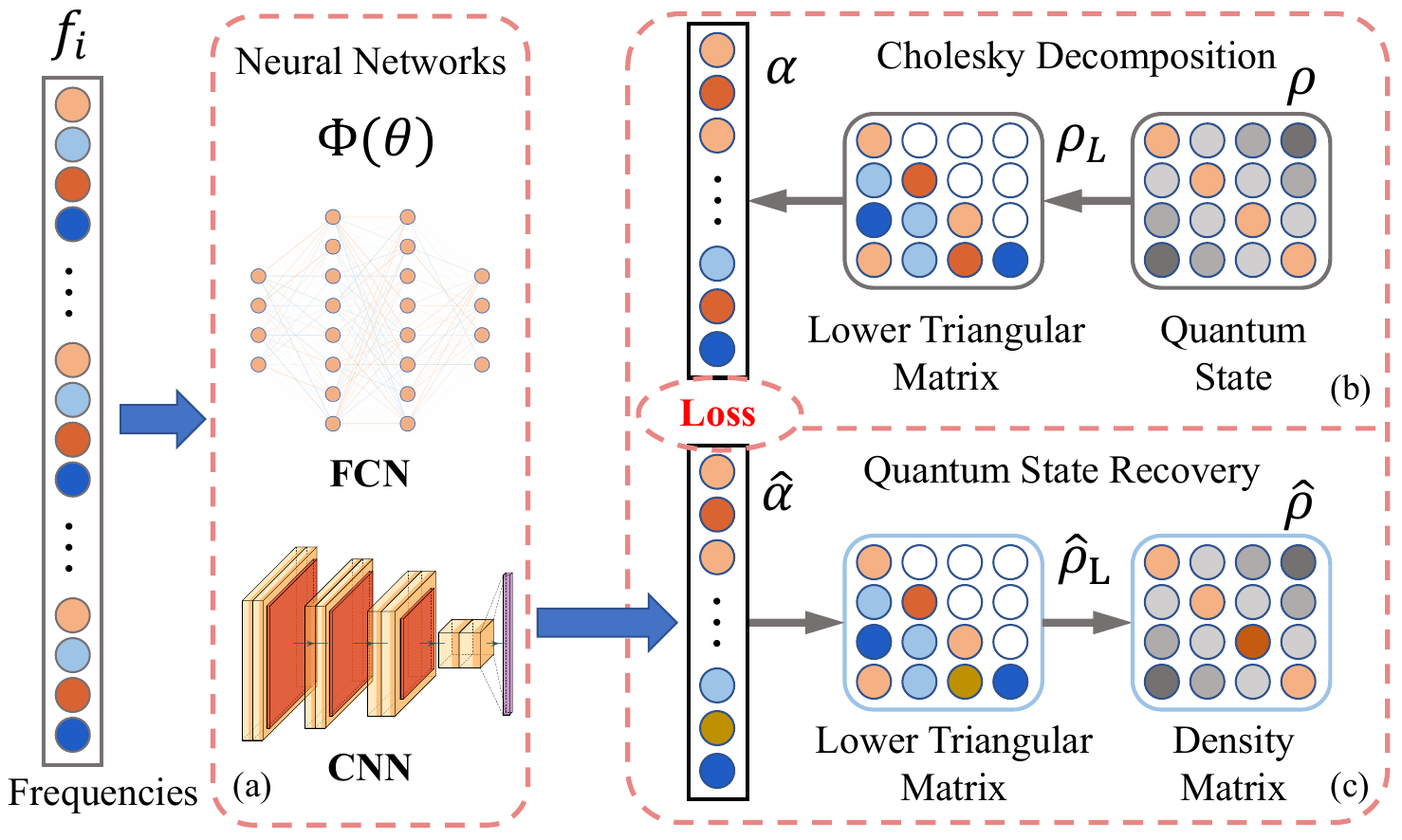}
	\caption{Schematic of the ML-based QST approach. (a) A multi-layer neural network $\Phi(\theta)$ maps the frequencies $\{f_i\}$ to the vector $\hat{\alpha}$; (b) Compute the Cholesky decomposition of the density matrices $\rho$ as the ground-truth vector $\alpha$ for supervised training with a loss function; (c) Obtain the quantum state $\hat{\rho}$ from the network's output $\hat{\alpha}$.}
	\label{fig:framework}
\end{figure}

QST involves extracting useful information from measured statistics \cite{jevzek2003quantum} and has been investigated using different neural networks. The reconstruction of density matrix $\rho$ from the measured frequencies $\{f_i\}$ can be simplified as searching for a complex function $\Phi$ that maps $\{f_i\}$ into $\rho$. Considering the physical features of a density matrix, we first consider how to generate a density matrix $\rho$ that is a positive semi-definite Hermitian operator with a unit trace. We may generate density matrices from lower triangular matrices
\begin{equation}
	\hat{\rho} = \frac{\rho_L\rho_L^{\dagger}}{\textup{Tr}(\rho_L\rho_L^{\dagger})}.
	\label{eq:tau2rho}
\end{equation}
In addition, according to the Cholesky decomposition~\cite{higham1990analysis}, for any physical density matrix, there exists a lower triangular matrix $\rho_L$ that achieves
\begin{equation}
	\rho_L\rho_L^{\dagger} = \rho.
	\label{eq:rho2tau}
\end{equation}

The recovery of a physical quantum state $\rho$ can be converted to retrieve a lower triangular matrix $\rho_L$ \cite{ahmed2021quantum}, which can be further transformed into a real vector $\alpha$ by splitting the real and imaginary parts. As such, QST using NNs is simplified as searching for a mapping function from the observed frequencies to the $\alpha$-vector that is closely associated with a density matrix. Though different architectures of NNs have been invented, they have a similar purpose of mapping the given input to the desired output, with a general framework presented in Fig.~\ref{fig:framework}. A multi-layer neural network is constructed to approximate a function mapping from measured frequencies $\{f_i\}$ to the intermediate vector $\hat{\alpha}$ which can be associated with a physical density matrix $\hat{\rho}$.

\begin{figure*}[ht]
	\centering
	\includegraphics[width=0.9\textwidth]{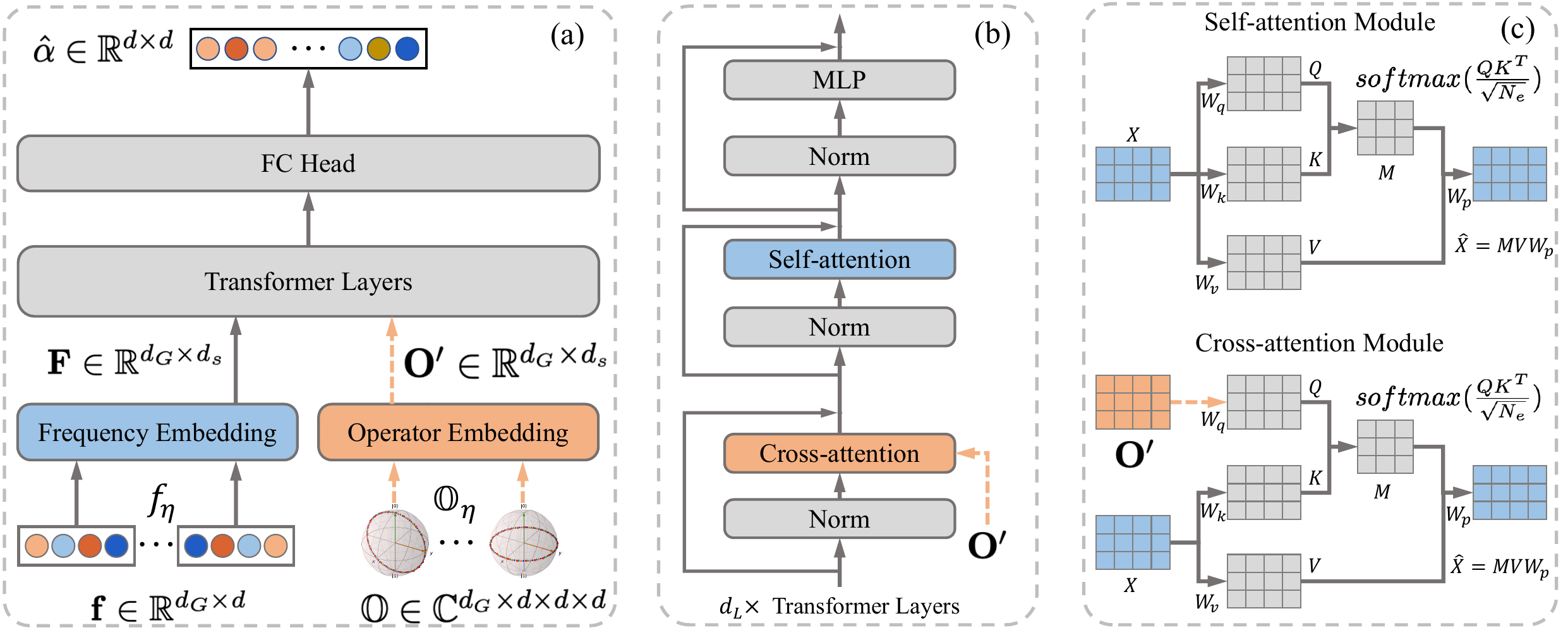}
	\caption{Framework for the proposed QAT-QST approach. (a) Two embedding modules are combined as the input for the network, namely frequency embedding, and operator embedding. Then, the transformer layers process the mainstream of frequency (Gray solid lines), and query the operator embedding (Orange dash lines), along with a fully connected head to obtain a real vector $\hat{\alpha}$. (b) The transformer layers consist of $d_L$ repeated stacks of normalization, a cross-attention module, normalization, a self-attention module, normalization, and multilayer perception layers. (c) The difference structures between self-attention and cross-attention module, where the cross-attention is used to learn how different elements in operator embedding are related to the mainstream.}
	\label{fig:transqst}
\end{figure*}

\section{Method}\label{Sec:method}

In this section, the framework of QAT-QST is first presented. Then the language structure in quantum measurements is analyzed, followed by the quantum-aware model design and quantum-aware loss function design. Finally, an integrated algorithm for the proposed method is summarized.

\subsection{Framework of QAT-QST}\label{sub:trans}

{\color{black}
	The quantum-aware transformer model is designed to leverage the structure of quantum measurements involved in QST, thereby translating experimentally measured frequencies into physical density matrices with high fidelity. The obtained data (i.e., a set of frequencies) is encoded as a set of vectors using language structures. Then, the realization of the quantum-aware transformer for QST is decomposed into two modules, namely frequency embedding to characterize density matrices, and operator embedding to allow for informative learning from additional information. They are combined and fed into the transformer layers to obtain the desired output. To enable the quantum-aware transformer to reconstruct quantum states with high fidelity, an integrated loss function that introduces the approximated Bures distance is proposed to guide the proposed QAT model to generate quantum states that have high fidelity with the original quantum states. 
	
	
	The schematic of QAT-QST is summarized in Fig. \ref{fig:transqst}. In Fig. \ref{fig:transqst}(a), the observed frequencies and the measurement operators are encoded into vectors that are fed into frequency embedding, and operator embedding, respectively. In Fig. \ref{fig:transqst}(b), the transformer layers along with a fully connected head are combined to obtain a real vector $\alpha$ that can be further utilized to generate a density matrix. In Fig. \ref{fig:transqst}(c), the transformer layers consist of $d_L$ repeated stacks of normalization, multi-head attention, normalization, and multilayer perception (MLP) layers. To guide the QAT model towards higher reconstruction fidelity, an integrated loss function that introduces an approximated Bures distance is designed to minimize the difference between the actual state and the retrieved state. 
	
	\subsection{Language structure in quantum measurements}
	
	In QST, a series of complete measurements are carried out to determine an unknown quantum state. These measurements can be represented by a set of measurement operators, denoted as $\mathbb{O}_{\eta}=\{\mathbb{O}_{\eta \gamma}\}$, where each POVM element $\mathbb{O}_{\eta \gamma}$ corresponds to a specific measurement outcome $\gamma$ in a detector \cite{lundeen2009tomography}.  For each detector $\mathbb{O}_\eta$, its elements must be positive semi-definite matrices that sum to the identity matrix, i.e., $\sum_ {\gamma}\mathbb{O}_{\eta \gamma}=\mathbb{I}$. This special feature of QST reminds us that the overall measurements can be grouped into different sets, each of which corresponds to a detector that contains multiple elements~\cite{wiseman2009quantum}. 
	
	In language modeling, several characters form one word, and several words constitute a full sentence. This structured representation of language modeling is very similar to that of QST using measurements. As illustrated in Fig. \ref{fig:lang}, each measurement corresponds to a character, and a group of measurements (referred to as a detector) corresponds to a word consisting of several characters. In quantum information fields, several groups of measurements specify a quantum state uniquely, similar to several words forming a sentence. In language modeling, the correlations might exist in characters among the same word or characters among different words. Similarly, in quantum scenarios, the correlations may come from two operators within one detector, or two operators among different detectors, which can be evaluated by matrix similarities. From that perspective, one detector in the overall measurements plays a similar role to a word in one sentence. 

	In language translation, words rather than characters are utilized as the basic elements for encoding~\cite{transformer2017}. This is in line with the fact that measurements in QST typically come in the form of detectors, which contain a set of operators that sum to an identity matrix. To solve QST effectively with structured measurements considered, the measured frequencies may be encoded in a language modeling manner. Let the number of operators in one detector be $d$ and let the number of detectors be $d_G$. In particular, the measured frequencies $\{f_i\}$ can be divided into $d_G$ groups, with each group containing $d$ elements. Denote the $\eta$-th group, $\eta\in \{1,2,...,d_G\}$ of measured frequencies as $\vec{f}_{\eta}=[f_1^\eta,f_2^\eta,...,f_{d}^\eta] \in\mathbb{R}^{1 \times d}$. $d_G$ groups of measured frequencies are finally concentrated into a two-dimensional matrix with the size of $d_G\times d$, where the $\eta$-th row denoting $\vec{f}_{\eta}$, playing a similar role to one word encoded as a context vector. One can refer to Appendix~\ref{app:example} for an example. Drawing from the language structure in QST, we utilize the transformer architecture to leverage the quantum patterns in measured frequencies, thereby achieving the task of QST by translating measured statistics into desired quantum states. 
	
	\subsection{Quantum-aware model design} 
	QST aims to reconstruct the density matrix of an unknown quantum state through a set of measurements. To efficiently model quantum states from experimental measurements, we utilize frequency embedding (the measured data from QST), and operator embedding (the additional information originated from quantum measurements), where embedding represents a relatively low-dimensional space into which one can translate high-dimensional vectors~\cite{transformer2017}. Finally, the combination of these vectors is input into the transformer layers equipped with multi-head attention modules to generate an $\alpha$ vector that is closely related to a physical density matrix.

	\textbf{Frequency embedding.}
	The vectors for structured measurements $\{\vec{f}_{\eta}\}_{\eta=1}^{d_G}$ compose a matrix $\mathbf{f}\in \mathbb{R}^{d_G \times d}$. Denote the dimension of the embedding size as $d_S$, and we have 
	\begin{equation}
		\mathbf{F} = \mathbf{f} \cdot \Theta_{f}, \quad \mathbf{F} \in \mathbb{R}^{d_G \times d_S}
		\label{eq:frequencyemb}
	\end{equation}
	where $\Theta_f\in \mathbb{R}^{d \times d_S}$ represents the weight of frequency embedding FCN layer, and $\cdot$ is the matrix multiplication. Similar to language modeling, the quantum state may be sensitive to the positional relations among individual qubits. For example, considering a 2-qubit measurement operator $\mathbb{O}_a=\mathbb{O}_1\otimes \mathbb{O}_2$ on a 2-qubit quantum state $\rho$, changing the order of $\mathbb{O}_1$ and $\mathbb{O}_2$ will likely have a completely different effect on $\rho$~\cite{zhong2022quantum}. To this end, we additionally utilize position embedding to abstract the relative or absolute position of frequencies measured by the operators. To guarantee that the position embedding has the same dimension as that of the frequency embedding, we use sine and cosine functions that are widely used in transformer implementation owing to their potential in learning relative position~\cite{brown2020gbt3,position}. 
	
	
	
	\textbf{Operator embedding.} Apart from the measured statistics, the measurement operators also provide significant information since traditional estimation methods involve the computation of the measurement operators. To achieve a high reconstruction fidelity, we integrate this valuable information into our model to extract an informative latent representation of quantum states. For each operator $\mathbb{O}_i\in \mathbb{C}^{d\times d}$, we split its real and imaginary parts to obtain a real vector (with length $2d^2$) and combine $d$ vectors (for $d$ operators in one detector) into a large vector (with length $2d^3$). Then, $d_G$ of large vectors (for $d_G$ detectors) compose a matrix as $\mathbf{O}\in \mathbb{R}^{d_G \times 2d^3}$. After this, an FC layer with training parameters $\Theta_m \in \mathbb{R}^{2d^3 \times d_S}$ projects $\mathbf{O}$ into $\mathbf{O'}$, calculated as: 
	\begin{equation}
		\mathbf{O'} = \mathbf{O} \cdot \Theta_{m},\quad \mathbf{O'} \in \mathbb{R}^{d_G \times d_S}.
		\label{eq:operatorembedding}
	\end{equation}
	
	\textbf{Quantum-aware transformer.}
	Now, the frequency embedding matrix $\mathbf{F}$ and operator embedding matrix $\mathbf{O'}$ share the same dimensions but exhibit distinct characteristics, so directly summarizing them as the input of a transformer network is an intuitive solution but not a satisfactory one.
	To address this, we reframe the QST problem as a multi-modal language task, which accommodates diverse input data types, including text and images, and leverages a cross-attention mechanism to extract and establish connections between information from these heterogeneous sources~\cite{radford2021learning,saharia2022photorealistic}.
	As a result, we anticipate that self-attention layers continue to integrate various types of information along the model's mainstream, enabling a holistic understanding of the frequency embedding matrix $\mathbf{F}$. Meanwhile, we expect the cross-attention layers to integrate the operator information into the mainstream latent representation by repeatedly querying the operator embedding matrix $\mathbf{O'}$.
	
	Let the input be $\mathbf{x} \in \mathbb{R}^{d_G \times d_S}$. In the self-attention module, $Q$, $K$ and $V$ follow the following formulation:
	\begin{equation}
		Q=\mathbf{x}\cdot{W_q}, \quad K=\mathbf{x}\cdot{W_k}, \quad V=\mathbf{x}\cdot{W_v},
		\label{eq:qkv-attn}
	\end{equation}
	where $W_q\in \mathbb{R}^{d_s \times N_e}$, $W_k\in \mathbb{R}^{d_s \times N_e}$, and $W_v\in \mathbb{R}^{d_s \times N_e}$ represent the weights of three FCN layers. In the cross-attention module, given the additional query input operator embedding matrix $\mathbf{O'}$, we have
	\begin{equation}
		Q=\mathbf{O'}\cdot{W_q}, \quad K=\mathbf{x}\cdot{W_k}, \quad V=\mathbf{x}\cdot{W_v},
		\label{eq:q-cross}
	\end{equation}
	The attention mechanism for both self-attention and cross-attention modules is uniformly formulated as
	\begin{equation}
		\hat{\mathbf{x}}=\textup{softmax} \left( \frac{Q \cdot K^T}{\sqrt{N_e}}\right)\cdot V \cdot W_p,\quad \hat{\mathbf{x}} \in \mathbb{R}^{d_G \times d_s},
		\label{eq:attention}
	\end{equation}
	where $N_e$ is the head number in the attention module and $W_p\in \mathbb{R}^{N_e \times d_s}$ represents the weight of an FCN layer to keep the input and output sharing the same dimensions. 
	Note that we haven't explicitly differentiated weights in Eq.~(\ref{eq:qkv-attn}) and Eq.~(\ref{eq:q-cross}), but each layer's self-attention and cross-attention modules have their own set of independent weights.
	
	The whole design of the quantum-aware transformer with $d_L$ layers is presented in Fig.~\ref{fig:transqst}. Recall that a positive Hermitian matrix $\rho$ is linked to a lower triangular matrix $\rho_L$, which can be represented as a real vector of length $d^2$, known as the $\alpha$-vector. To obtain this representation, a fully connected head layer is applied to convert the transformer output into an $\hat{\alpha}$-vector of size $d^2$. This $\hat{\alpha}$-vector can be utilized to generate a corresponding physical density matrix through additional operations (refer to Fig.~\ref{fig:framework} for detailed information). 

	\subsection{Quantum-aware loss function}
	\textbf{Bures distance.} In quantum information theory, there are several choices for evaluating the similarity of quantum states. Given two quantum states $\rho_1$ and $\rho_2$, their fidelity is defined as 
	\begin{equation}
		\mathcal{F}(\rho_1,\rho_2)=(\textup{Tr}\sqrt{\sqrt{\rho_1},\rho_2,\sqrt{\rho_1}})^2.
		\label{eq:fidelity}
	\end{equation}
	Furthermore, we have their \emph{Bures} distance~\cite{marian2008bures}, given by
	\begin{equation}
		D_B(\rho_1,\rho_2)=2(1-\sqrt{\mathcal{F}(\rho_1,\rho_2)}).
		\label{eq:bures distance}
	\end{equation}
	\emph{Angle} metric is given as 
	\begin{equation}
		D_A(\rho_1,\rho_2)= \arccos \sqrt{\mathcal{F}(\rho_1,\rho_2)}.
		\label{eq:angle}
	\end{equation}
	Considering the specific features of quantum states, it is highly desirable to design a novel loss function that accounts for the ``closeness" (the fidelity) of two quantum states. 
	
	However, the fidelity function in Eq.~(\ref{eq:fidelity}) involves computations of eigenvalues of complex matrices, which brings high overhead when training the neural networks with a back-propagation process. Meanwhile, reconstructing a quantum state with density matrix $\rho$ has been converted to retrieve a real vector $\hat{\alpha}$, which is in one-to-one correspondence to $\rho$. Hence, we try to approximate the Bures distance concerning a real vector in the Euclidean space. From Eq.~(\ref{eq:bures distance}) and Eq.~(\ref{eq:angle}), we have $D_B=2(1-\cos (D_A))$, with $D_A$ denoting the angle between two quantum states. For two vectors $\alpha$ and $\hat{\alpha}$, there exists an angle $\theta_{v}$ that might provide implications for evaluating quantum fidelity. For example, we may have the approximated Bures distance in the Euclidean space as $\tilde{D}_B(\boldsymbol{\alpha},\boldsymbol{\hat{\alpha}})=(1-\cos(\theta_{v}))$ (neglecting the ratio of 2). Intuitively, it shares a similar formulation to the cosine distance, $(1-\cos(\theta_v))$, where $\cos(\theta_v)$ corresponds to the cosine similarity in the ML community.

	
	\begin{figure}[ht]
		\centering
		\includegraphics[width=0.40\textwidth]{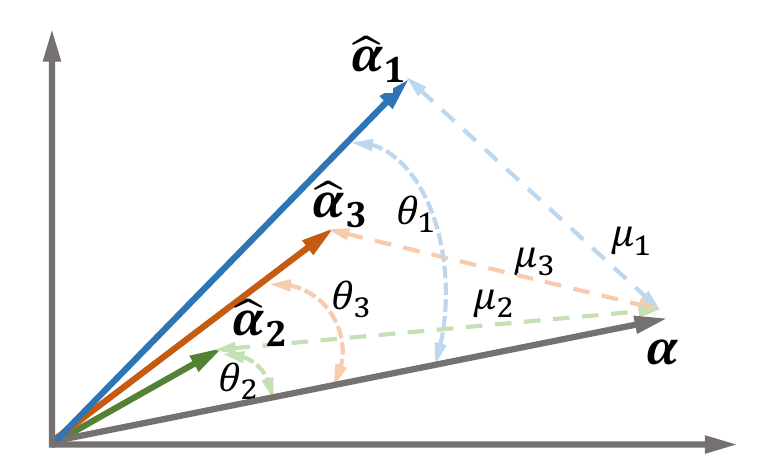}
		\caption{Visualization of three vectors in different criteria.  $\boldsymbol{\alpha}$ represents the fixed target vector from the Cholesky decomposition.  $\boldsymbol{\hat{\alpha}_1}$, $\boldsymbol{\hat{\alpha}_2}$, $\boldsymbol{\hat{\alpha}_3}$ represent three candidate vectors. If one needs to choose one vector close to $\boldsymbol{\alpha}$, $\boldsymbol{\hat{\alpha}_1}$ is a good candidate regarding the Euclidean distance because $\mu_1 < \mu_2$; $\boldsymbol{\hat{\alpha}_2}$ is a good candidate 
				regarding the Bures distance because $\theta_2 < \theta_1$. When considering both the Bures distance and the  Euclidean distance with the target vector $\boldsymbol{\hat{\alpha}}$, $\boldsymbol{\hat{\alpha}_3}$ is a good candidate that achieves good performance on both criteria. As such, $\boldsymbol{\hat{\alpha}_3}$ is a more suitable candidate vector.
		}
		\label{fig:lossfunction}
	\end{figure}
	
	\textbf{Qualitative analysis.}
	Based on the above observations, utilizing the angle of $\alpha$-vectors in the Euclidean space introduces an approximated Bures distance. We should still consider the Euclidean distance that accounts for the amplitude of quantum states. Denote the approximated Bures distance as $\upsilon(\boldsymbol{\alpha},\boldsymbol{\hat{\alpha}})$ with the angle as $\theta=\arccos{\upsilon}(\boldsymbol{\alpha},\boldsymbol{\hat{\alpha}})$. Denote the Euclidean distance as $\mu(\boldsymbol{\alpha},\boldsymbol{\hat{\alpha}})$. We may clarify their separate impacts in one example. As is illustrated in Fig. \ref{fig:lossfunction}, given a target vector $\boldsymbol{\alpha}$, there are three candidate vectors, i.e., $\boldsymbol{\hat{\alpha}_1}$, $\boldsymbol{\hat{\alpha}_2}$, $\boldsymbol{\hat{\alpha}_3}$ that is generated from the NNs. One needs to choose one vector that is close to $\boldsymbol{\alpha}$. 
		$\mu_1 < \mu_2$ suggests that $\boldsymbol{\hat{\alpha}_1}$ is a good candidate with low Euclidean distance against the target vector $\boldsymbol{\alpha}$. $\cos(\theta_2) > \cos(\theta_1)$ and $\theta_2 < \theta_1$ reveals that 
		$\boldsymbol{\hat{\alpha}_2}$ is a good candidate with low Bures distance against the target vector $\boldsymbol{\alpha}$. With both the two criteria considered, $\boldsymbol{\hat{\alpha}_3}$ is the best candidate vector that is close to the target vector $\boldsymbol{\alpha}$, owing to $ \mu_2 < \mu_3 < \mu_1$ and $\cos(\theta_1) < \cos(\theta_3) < \cos(\theta_2)$ with $\theta_2 < \theta_3 < \theta_1 $. 
	
	\textbf{Integrated quantum-aware loss.}
	To enable the QST model to generate the $\alpha$ vector that brings in high similarity with quantum states, we define a new loss function that incorporates the approximated Bures distance into the Euclidean distance as
	\begin{equation}
		\mathbb{L}(\boldsymbol{\alpha},\boldsymbol{\hat{\alpha}})= \beta \upsilon(\boldsymbol{\alpha},\boldsymbol{\hat{\alpha}})+(1-\beta)\mu(\boldsymbol{\alpha},\boldsymbol{\hat{\alpha}}) ,
		\label{eq:unified}
	\end{equation}
	where $\beta$ is a hyper-parameter balancing the trade-off between the two factors. By incorporating the approximated Bures distance (narrowing the Bures angle with high state fidelity) into the traditional MSE criterion (converging to the target value with small amplitude errors), this integrated function is useful to guide the quantum-aware transformer for QST in reconstructing quantum states with high fidelity.

	\emph{Remark 1.} In the ML community, MSE is a loss function that measures the average squared distance between the ground-truth vectors and the predicted vectors. MSE and Euclidean distance share a similar characteristic of measuring the distance between two vectors, with MSE being the square of Euclidean distance. In this work, we use the notation of Euclidean distance to maintain consistency with approximated Bures distance and to emphasize the physical interpretation of the metric between two vectors. But, the exact value regarding the Euclidean criterion is calculated using the MSE loss.

	\subsection{Integrated algorithm}\label{subsec:algorithm}
	\begin{algorithm}[ht]
		\caption{Algorithm description for QAT-QST}\label{al:AF-QST}
		\KwData{Measurement operators $\{\mathbb{O}_{\eta}\}$ and measured frequencies $\{\vec{f}_{\eta}\}$}
		\KwResult{The reconstructed states $\hat{\rho}$}
		
		\emph {Build a QAT-QST model, i.e, Eq.~(\ref{eq:mapfunction})}\; 
		\quad Implement the frequency and operator embeddings following Eq.~(\ref{eq:frequencyemb}) and Eq.~(\ref{eq:operatorembedding}), respectively\;
		\quad Implement the self-attention module on $\mathbf{x}$ following Eq.~(\ref{eq:qkv-attn}) and Eq.~(\ref{eq:attention})\;
		\quad Implement the cross-attention module on $\mathbf{O}^{\prime}$ following Eq.~(\ref{eq:q-cross}) and Eq.~(\ref{eq:attention})\;
		\quad Perform $d_L$ transformer layers\;
		\quad Obtain $\alpha$-vector through an FC network\;
		\emph{Optimize the QAT-QST model}\;
		\While{not convergent}{
			Obtain the output from the Cholesky decomposition as ground truth for $\alpha$-vector\;
			
			Obtain NNs' output as the actual value for $\alpha$-vector\;
			Measure integrated loss function using Eq.~(\ref{eq:unified})\;
			Update the parameters using the gradient descent method\;
		}
		\emph {Test the QAT-QST model}\;
		\quad Obtain the $\alpha$-vector using Eq.~(\ref{eq:mapfunction})\;
		\quad Obtain the physical density matrix using Eq.~(\ref{eq:tau2rho}).
		
	\end{algorithm}
	
	Given the observed frequencies and the associated measurement operator, reconstructing quantum states from measured statistics can be summarized as a parameterized function, formulated as 
	\begin{equation}
		{\hat{\alpha}} = \Phi({\Theta};\{\vec{f}_\eta\},\{\mathbb{O}_{\eta}\}),
		\label{eq:mapfunction}
	\end{equation}
	where $\Theta$ are all the trainable parameters involved in the NN blocks in Fig. \ref{fig:transqst}. The integrated procedure is summarized in Algorithm~\ref{al:AF-QST}. Given the measured frequencies and the measurement operators, the frequency embedding and the operator embedding are implemented. Then, their combination is transmitted to the transformer, which contains $d_L$ repeated stacks of normalization, multi-head attention, normalization, and MLP layers. Through an FC layer, a real vector $\hat{\alpha}$ is obtained that can further produce a physical density matrix according to Eq.~(\ref{eq:tau2rho}).
	

	\section{Results}\label{Sec:results}
	This section first presents the implementation details of the proposed QAT-QST approach and other QST methods, followed by the main results of reconstructing pure and mixed states. Investigation of the separate impacts of the quantum-aware model design and quantum-aware loss function design is performed. Finally, experimental results on IBM computers are provided. 
	
	\subsection{Implementation details}\label{subsec:setting}
	\textbf{Dataset settings.}
	In this work, we focus on QST for both pure and mixed quantum states. The generation of random pure states can be achieved through the use of the Haar metric~\cite{mezzadri2006generate}. Let $\mathbb{U}^d$ represent the set of all $d$-dimensional unitary operators, meaning that any $U \in \mathbb{U}^d$ satisfies $UU^{\dagger} =U^{\dagger}U = \mathbb{I}_d$. The Haar metric is actually a probability measure on a compact group due to its invariance under group multiplication~\cite{mezzadri2006generate}. This property enables generating unitary transformations using the Haar measure~\cite{danaci2021machine}, leading to the generation of random pure states as 
	\begin{equation}
		\rho_{haar} =  U_{Haar} |\psi_0\rangle \langle \psi_0| U_{Haar}^{\dagger },
		\label{eq:haar}
	\end{equation}
	where $|\psi_0\rangle$ is a fixed pure state. For mixed states, we consider the random matrix from the Ginibre ensembles given by~\cite{forrester2007eigenvalue} 
	\begin{equation}
		G=\mathcal{N}_d(0,1)+\textup{i}\mathcal{N}_d(0,1),
		\label{eq:ginibre ensemble}
	\end{equation}
	where $\mathcal{N}_d(0,1)$ represents the random normal distributions of size $d \times d$ with zero mean and unity variance. Random density matrices using the Hilbert-Schmidt metric are given by~\cite{ozawa2000entanglement}
	\begin{equation}
		\rho_{ginibre}=\frac{GG^{\dagger}}{\textup{Tr}(GG^{\dagger})}.
		\label{eq:ginibre state}
	\end{equation}
	
	For measurements, we first consider tensor products of Pauli matrices, which is also called cube measurement~\cite{de2008choice}. Denote the Pauli matrices as $\sigma=(\sigma_x,\sigma_y,\sigma_z)$, with
	\begin{equation}
		\sigma_x=\left[\begin{array}{cc}
			1 & 0\\
			0 & -1
		\end{array}\right],
		\sigma_y=\left[\begin{array}{cc}
			0 & -\rm{i}\\
			\rm{i} & 0
		\end{array}\right],
		\sigma_z=\left[\begin{array}{cc}
			0 & 1\\
			1 & 0
		\end{array}\right].
	\end{equation}
	Let $|H\rangle$ and $|V\rangle$ be eigen-vectors of  $\sigma_z$, $|L\rangle$ and $|R\rangle$  be eigen-vectors of $\sigma_y$, and $|D\rangle$ and $|A\rangle$ be eigen-vectors of  $\sigma_x$. For $n$-qubit cube measurement, $3^n$ detectors are involved, with each detector containing $d=2^n$ measurement operators~\cite{adamson2010improving}. For example, for a 2-qubit case,  $|HH\rangle$,$|HV\rangle$,$|VH\rangle$, $|VV\rangle$ correspond to a real detector that can be experimentally realized on quantum devices. Apart from that, we also investigate different measurement operators that can be assigned to different quantum states. Considering square root measurement has great potential for distinguishing quantum states~\cite{motka2017efficient}, we construct a candidate pool composed of different square root measurements. In particular, a set of $d$ square root measurements compose a detector. For each operator $\gamma\in\{1,2,...,d\}$, we have 
	\begin{equation}
		\mathbb{O}_{\gamma} = \Lambda ^{-1/2}|\psi_\gamma\rangle \langle \psi_\gamma| \Lambda^{-1/2},\quad \textup{with} \quad \Lambda =\sum_{\gamma} |\psi_{\gamma} \rangle \langle \psi_{\gamma}|,
		\label{eq:srm}
	\end{equation}
	where $\{|\psi_{\gamma}\rangle\}$ are randomly generated Haar-distributed pure states~\cite{mezzadri2006generate}. During the measurement process, the number of total copies for each detector is set as $10000$ in this work. 
	
	\textbf{Training and testing settings.}
	In this work, 100,000 states are randomly generated by sampling the parameters in Eq.~(\ref{eq:haar}) and Eq.~(\ref{eq:ginibre ensemble}) for pure and mixed states on 2/3/4-qubit systems, respectively. 95,000 states are used for training the QAT-QST model, while the remaining 5,000 states are left for evaluation. Here, the measured frequencies in Eq.~(\ref{eq:frequency}) used for training and testing are not error-free, as they suffer from noise due to the deviation from the true probabilities in Eq.~(\ref{eq:prob}). Typically, a large number of copies is sufficient to ensure that the measured frequencies do not deviate significantly from actual probabilities in practical applications. However, in our simulations, experiments with limited copies are intentionally implemented to investigate the performance of the proposed method in dealing with simulated noisy data. 
	
	
	The performance of a typical machine learning task with an affirmative schematic is largely determined by the training strategies and model complexities~\cite{lecun2015deep}. To address this, we conduct a comprehensive simulation to assess the impact of various training strategies and evaluate the performance of different models (detailed information in Appendix ~\ref{subsec:model}). The training strategies for all cases are as follows: we use the Adam optimizer with default settings, a batch size of 256, an initial learning rate of $0.005$ with cosine learning rate attenuation, and 500 training epochs with a warm-up strategy for the first 20 epochs, during which the learning rate is gradually increased from 0 to $0.005$. After exploring various combinations of depth and width in the models (detailed in Appendix \ref{subsec:model}), we choose parameter settings as $d_L=8$, $d_S=32$, $d_H=16$, and $ d_{rate}=8$.
	
	After the NNs' model is learned, we directly apply it to the testing samples to obtain the $\hat{\alpha}$ vectors, which are utilized to generate the final density matrices using Eq.~(\ref{eq:tau2rho}). To demonstrate the efficiency of QST, the fidelity between the reconstructed state $\hat{\rho}$ and the real state $\rho$, i.e., $\mathcal{F}(\rho,\hat{\rho})$~is measured. Additionally, we can measure the infidelity as $1-\mathcal{F}(\rho,\hat{\rho})$ and log of infidelity as $\log(1-\mathcal{F}(\rho,\hat{\rho}))$.
	
	\textbf{Parameter settings for other QST methods.} In~\cite{ma2024neural}, it was found that maximum likelihood estimation (MLE)~\cite{jevzek2003quantum} (the most popular traditional tomography algorithm) generally performs similarly to linear regression estimation (LRE)~\cite{qi2013quantum,xiao2024quantum} for pure states and has a slight advantage over LRE for mixed states. However, since LRE is more computationally efficient than MLE, we include it as the traditional method for comparison. To focus on the capability of our method, we also implement other two ML-based methods, including FCN-based QST~\cite{ma2024neural} and CNN-based QST~\cite{lohani2020cnn}, which have been studied in recent decades.
	FCN-based QST utilizes 5 hidden layers, with each hidden layer containing 256 neurons. The general architecture of CNN is based on the design in ~\cite{lohani2020cnn}. To enhance the expressiveness, a kernel size of $3\times3$ is employed in the two-dimensional convolutional layers. The first and second layers have channel sizes of 64 and 128, respectively. During the training process, the Adam optimization with $\beta1=0.9$ and $\beta2=0.999$ is utilized, with a batch size of 256 and 500 training epochs, including a warm-up phase for the initial 20 epochs. It is worth noting that the initial learning rate is set as $0.0001$ for FCN and $0.001$ for CNN.

	\begin{figure*}[ht]
		\centering	
		\subfigure[Pure states.]
		{\includegraphics[width=0.45\linewidth]{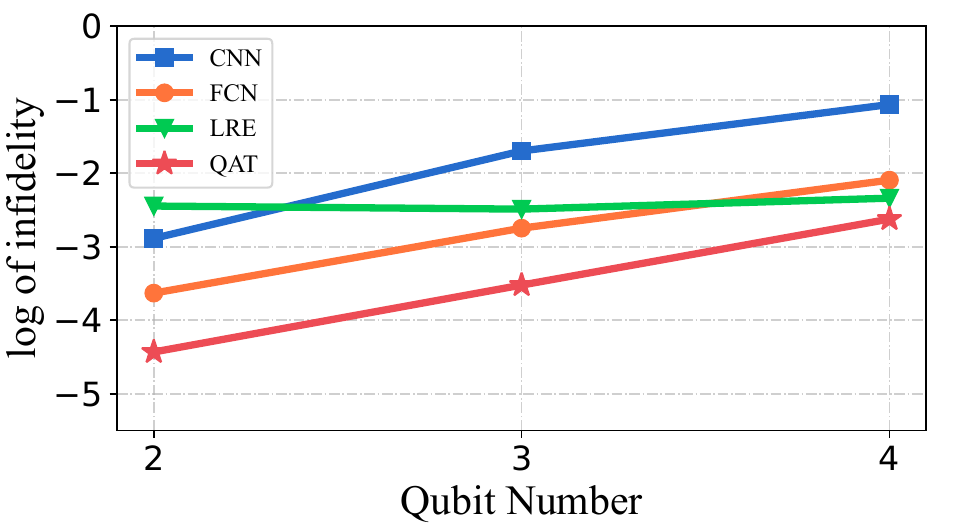}} \label{sfig:flops}
		\subfigure[Mixed states.]
		{\includegraphics[width=0.45\linewidth]{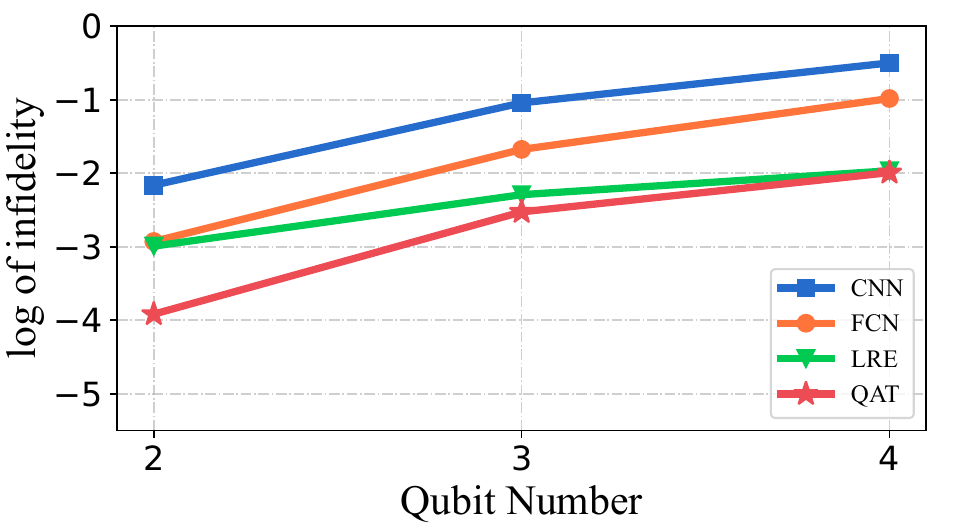}} \label{sfig:speed}
		\caption{The comparison of QST using different methods.}
		\label{fig:sota}	
	\end{figure*}
	
	\begin{figure}[ht]
		\centering	
		\includegraphics[width=0.95\linewidth]{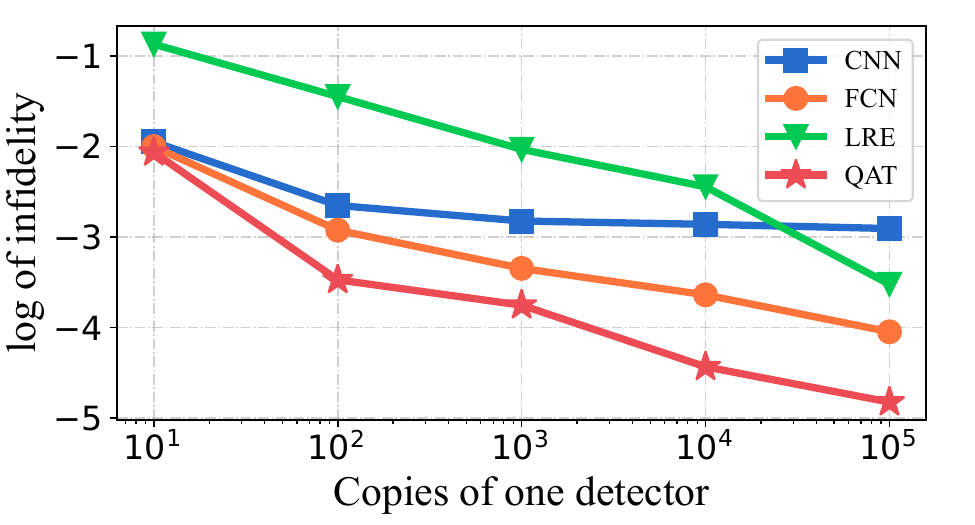} 
		\caption{The performance of 2-qubit tomography using different copies. }
		\label{fig:copy}	
	\end{figure}
	
	\subsection{Main results}\label{subsec:main}
	
	
	Due to the exponential scaling issues in full QST, we implement simulations on low qubits in this work. To verify the effectiveness of the proposed method,  QST of 2/3/4-qubit states is conducted on both pure states and mixed states. The numerical results are summarized in Fig.~\ref{fig:sota}, where the proposed method using the quantum-aware transformer outperforms the other methods when reconstructing pure states, with CNN achieving the lowest performance. For mixed states, the proposed method achieves the highest results on 2 qubits and exhibits similar performance to LRE on 3/4 qubits. Additionally, the gaps between the four methods for pure states are larger than those for mixed states.

	It is known that deep NNs can approximate different patterns of data efficiently with favorable generalization properties~\cite{lecun2015deep} and  QST using NNs, e.g., FCN \cite{ma2024neural}, has exhibited potential for limited copies, which is usually the case in practical applications. Therefore, we also explore the favorabilities of QAT-QST using a few measurement copies. The numerical results are summarized in Fig. \ref{fig:copy}, where the log of infidelity for the proposed method decreases with the measurement resources, showing the same trend as LRE. However, there exist great gaps between QAT-QST and LRE, which demonstrates the superiority of the proposed method in dealing with limited resources. Low values for measurement copies result in a gap between the measured frequencies and the true probabilities, causing noise in measured data. From this perspective, the good performance of QAT-QST under different numbers of copies with superiority over other methods verifies the robustness of the proposed model in dealing with noise in the data.  
	
	\begin{figure*}[]
		\centering	
		\subfigure[Pure states]
		{\includegraphics[width=0.4\linewidth]{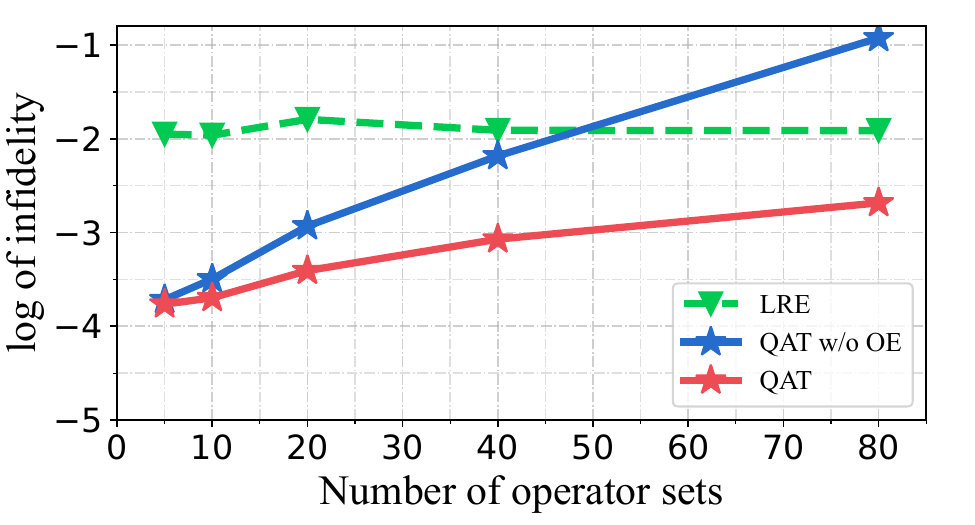} \label{sfig:psqrm}}
		\subfigure[Mixed states]
		{\includegraphics[width=0.4\linewidth]{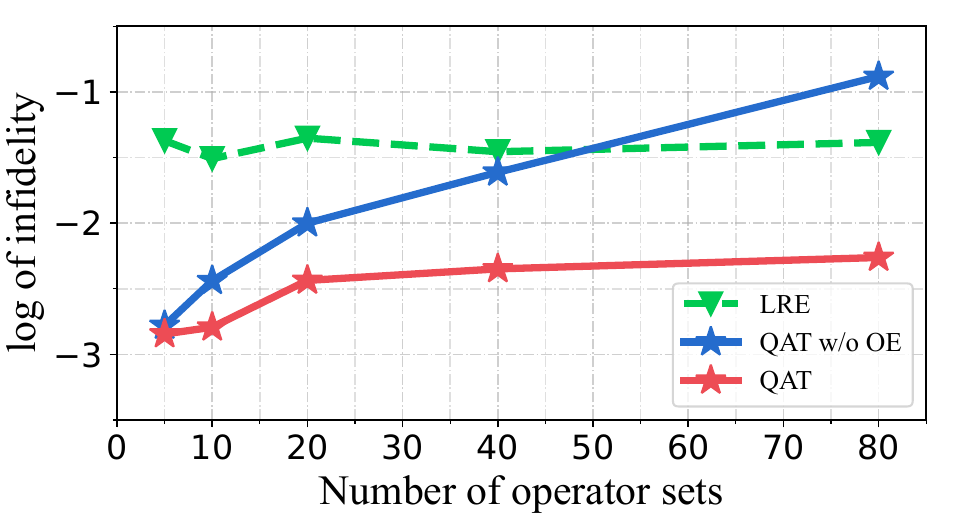} \label{sfig:msqrm}}
		\caption{QST using square root measurements with different numbers of operator sets. OE: operator embedding.}
		\label{fig:operator-embedding}	
	\end{figure*}
	
	\begin{figure*}[]
		\centering
		\subfigure[Train with MSE loss]
		{\includegraphics[width=0.3\linewidth]{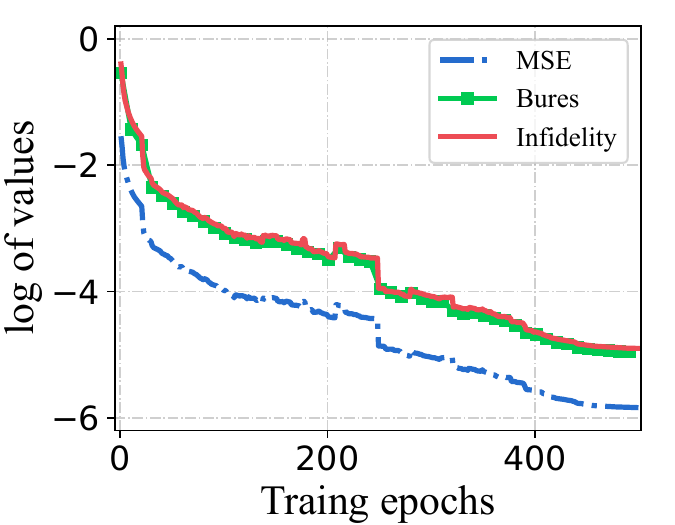} \label{sfig:flops}}
		\subfigure[Train with Bures distance loss]
		{\includegraphics[width=0.3\linewidth]{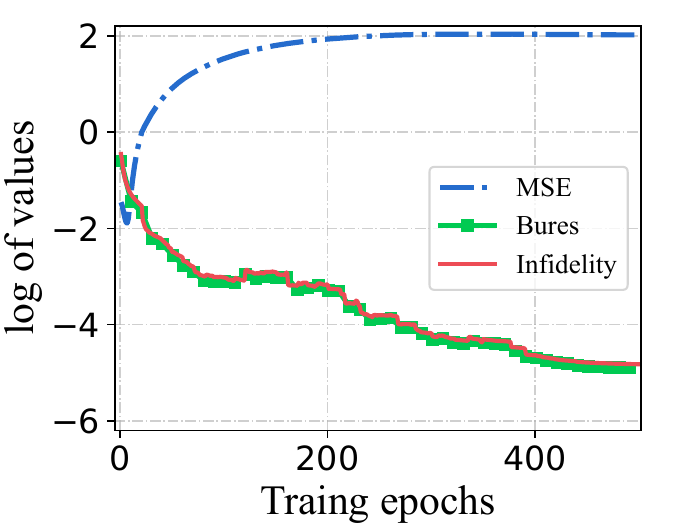} \label{sfig:flops}}
		\subfigure[Train with Integrated loss with $\beta=0.09$]
		{\includegraphics[width=0.3\linewidth]{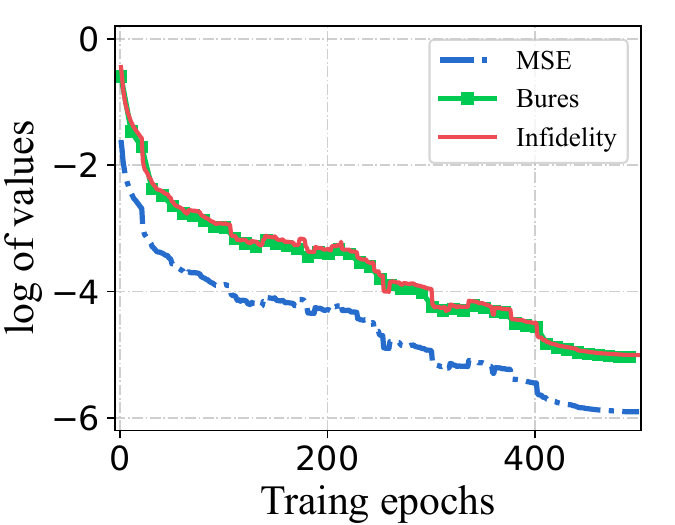} \label{sfig:speed}}	
		\caption{The learning curves of QAT under different loss functions. Each figure is trained with a specific loss function, but the actual values of the MSE distance, the approximated Bures, and infidelity are also calculated to investigate their relationships. (a) under the MSE loss, infidelity (Infid) converges to $1.25\times10^{-5}$, (b) under the Bures distance loss, infidelity converges to $1.50\times 10^{-5}$, (c) under the integrated loss,  infidelity converges to $9.81\times 10^{-6}$.}
		\label{fig:embedloss}	
	\end{figure*}

	
	\begin{table*}[]
		\centering
		\caption{Infidelity of testing states when the QAT model is trained using different loss functions.}
		\begin{tabular}{c|ccc|ccc}
			\midrule[1.5pt]
			& \multicolumn{3}{c|}{Pure states} & \multicolumn{3}{c}{Mixed states} \\
			\midrule[1.5pt]
			Qubit number & 2      & 3      & 4   & 2      & 3      & 4  \\
			\midrule[1.5pt]
			MSE        & $1.25\times10^{-5}$ & $6.62\times10^{-5}$ & $1.98\times10^{-4}$ & $1.44\times10^{-4}$ & $6.37\times10^{-3}$ & $9.29\times10^{-2}$\\
			Bures distance        & $1.61\times10^{-5}$ & $4.90\times10^{-5}$ & $1.93\times10^{-4}$ & $1.31\times10^{-4}$ & $3.53\times10^{-3}$ & $7.19\times10^{-2}$\\
			Integrated    & $9.81\times10^{-6}$ & $4.84\times10^{-5}$ & $1.75\times10^{-4}$& $1.17\times10^{-4}$ & $3.03\times10^{-3}$ & $4.72\times10^{-2}$\\
			\toprule[2pt]
		\end{tabular}
		\label{tabel:loss}
	\end{table*}
	
	
	\begin{table*}[]
		\centering
		\caption{Fidelity of reconstructing quantum states on  $\emph{ibmq\_{manila}}$.}
		\begin{tabular}{c|ccc|ccc}
			\midrule[1.5pt]
			& \multicolumn{3}{c|}{100 shots } & \multicolumn{3}{c}{1000 shots} \\
			\midrule[1.5pt]
			& QAT    &  LRE  & IBM built-in & QAT    &  LRE  & IBM built-in  \\
			\midrule[1.5pt]
			Mean     & 0.975340 & 0.904665 & 0.895242 & 0.994796 & 0.917861 & 0.917274 \\
			Min      & 0.918696 & 0.822451 & 0.803072 & 0.975933 & 0.820536 & 0.819485 \\
			Max      & 0.995288 & 0.961351 & 0.957352 & 0.999379 & 0.972831 & 0.965420 \\
			Variance & 3.02$\times10^{-4}$ & 1.02$\times10^{-3}$ & 1.11$\times10^{-3}$ & 1.77$\times10^{-5}$ & 7.92$\times10^{-4}$ & 7.96$\times10^{-4}$ \\
			\toprule[2pt]
		\end{tabular}
		\label{tabel:experiment_manila}
	\end{table*}
	
	\begin{table*}[]
		\centering
		\caption{Fidelity of reconstructing quantum states on  $\emph{ibmq\_{belem}}$.}
		\begin{tabular}{c|ccc|ccc}
			\midrule[1.5pt]
			& \multicolumn{3}{c|}{100 shots} & \multicolumn{3}{c}{1000 shots} \\
			\midrule[1.5pt]
			& QAT    &  LRE  & IBM built-in & QAT    &  LRE  & IBM built-in\\
			\midrule[1.5pt]
			Mean     & 0.976820 & 0.899702 & 0.890585 & 0.993845 & 0.913588 & 0.911888 \\
			Min      & 0.919605 & 0.828027 & 0.807011 & 0.984138 & 0.866645 & 0.860420  \\
			Max      & 0.996530 & 0.982373 & 0.973111 & 0.998958 & 0.963624 & 0.959016 \\
			Variance & 2.46$\times10^{-4}$ & 1.25$\times10^{-3}$ & 1.35$\times10^{-3}$ & 1.14$\times10^{-5}$ & 6.06$\times10^{-4}$ & 6.47$\times10^{-4}$ \\
			\toprule[2pt]
		\end{tabular}
		\label{tabel:experiment_belem}
	\end{table*}
	
	\subsection{Investigation of quantum-aware design}\label{subsec:ab}
	Apart from the language structure involved in quantum measurements that fit into the design of the transformer, we introduce two key elements to build a quantum-aware transformer for QST, including (i) querying quantum operators to allow for informative representation learning of quantum states, and (ii) employing Bures distance to realize precise evaluation of quantum state fidelity. To clarify their significance on QST, extensive comparisons regarding the quantum-aware model design (the operator embedding) and the quantum-aware loss design (i.e., the Bures distance) are implemented. 
	
	
	\textbf{Operator embedding for quantum-aware model design.} To clarify the significance of operator embedding, we focus on square root measurements, i.e., quantum states are measured using different measurements rather than fixed CUBE measurements. To specify quantum states using informationally complete measurements, five detectors among the above candidate pool in Eq.~(\ref{eq:srm}) are sampled to compose the overall measurements for QST. When $M=5$, all states utilize the same measurement settings. A large $M$ implies that states are measured using different measurement settings, meaning the samples cover a broader range of space, making it more challenging to accommodate a generalized QST method capable of handling various measurement operators. As illustrated in  Fig.~\ref{fig:operator-embedding}, the improvement of the operator embedding increases with $M$, demonstrating that the necessity to utilize the operator embedding is strong when observing quantum states using different measurements.
	
	
	
	\textbf{Bures distance for quantum-aware loss design.} From (\ref{eq:unified}), the integration of Bures distance in the loss function can be controlled by the value of $\beta$. Before determining a good value of $\beta$, we first train the QAT-QST model with MSE loss~\cite{ma2024neural} on 2-qubit pure states and find that the value of the approximated Bures distance is about 10 times higher than the MSE value. Hence, we choose the hyper-parameter $\beta=0.09$ to ensure the consistency of the magnitude of the gradient values of the two parts when conducting the backpropagation process. To verify the impacts of different $\beta$, three sets of values for $\beta$ are compared, with $\beta=1$ (MSE loss or Euclidean distance), and $\beta=0$ (approximated Bures distance) and $\beta=0.09$ (a combination of Bures distance and Euclidean distance). The learning curves for reconstructing 2-qubit pure states under different loss functions are summarized in Fig.~\ref{fig:embedloss}. In each subfigure, one loss function with fixed $\beta$ is chosen to train the parameters of QAT-QST, but the remaining distance criteria together with the log of infidelity are also measured to clarify their relationships. The log of infidelity exhibits similar values to that of approximated Bures distance, demonstrating the effectiveness of approximated Bures distance in evaluating the fidelity of quantum states. In Fig. ~\ref{fig:embedloss}(b), when only taking the approximated Bures distance as the loss function, the MSE can be extremely large since magnitude is not penalized. Hence, a good solution is to integrate the Bure distance into the conventional MSE loss to guide the training of the QAT model with high fidelity.

	
	After the training process is completed, we also implement QST on testing samples and summarize the infidelity value in Table~\ref{tabel:loss}. For pure states, the approximated Bures distance achieves a comparative performance to that of the MSE loss. Notably, the integrated loss achieves the highest results. When reconstructing mixed states, the approximated Bures distance enhances the performance compared with MSE loss. Correspondingly, the integrated loss brings a big improvement over the MSE loss. The results demonstrate that the integrated loss integrates the benefits of Euclidean and approximated Bures distances to enable a better training performance for QAT-QST.
	


	
	\subsection{Experimental resulst}
	
	To demonstrate the potential of the proposed QAT-QST approach, we perform QST experiments on IBM quantum computers. Due to the high cost of obtaining resources on IBM devices, we performed CUBE measurements on 100 states for testing.  The experiments on IBM quantum circuits are executed as follows: Initialize quantum circuits via \emph{QuantumCircuit.initialize} from the Qiskit Terra API) to prepare quantum states, denoted as ideal reference states. The uploaded circuits are further transpiled by the backend. We then execute the transpiled circuits for 100 and 1000 shots on different quantum devices (including $\emph{ibmq\_{manila}}$ and $\emph{ibmq\_{belem}}$) and record the measurement counts, respectively. Here, the number of shots in IBM quantum devices corresponds to the number of total copies for one detector.

	The QAT model is trained using the simulated data in Subsection~\ref{subsec:setting}, where the states for training and the states for testing are actually from the same distribution, and their measured statistics were obtained with the same number of shots, which is the same as the experimental setting, i.e., 100 shots and 1000 shots. For experiments, we prepare quantum states that belong to the testing dataset and obtain measured statistics for further reconstruction. The evaluation is based on the fidelity between the density matrix of the ideal reference state and the density matrix reconstructed from experimentally measured statistics. The comparison results of performing QST on the two devices in Table~\ref{tabel:experiment_manila} and Table~\ref{tabel:experiment_belem} demonstrate the favorable robustness of our method in dealing with experimental data, compared with conventional methods (LRE and the IBM built-in method). The comparsion strengthens the capability of our proposed method in practical quantum applications. The notable improvement primarily stems from the fact that NN-based QST endeavors to approximate a mapping function from measured statistics to density matrices. Hence, the trained model has generalization in testing samples and exhibits robustness against errors.
	
	
	\section{Conclusion}\label{Sec:conclusion}
	
	In this paper, we investigated the intrinsic patterns in quantum measurements for QST and found that a detector (composed of a set of measurements with sum to identity) plays a similar role to words in one sentence. To leverage the quantum patterns in QST, we proposed a quantum-aware transformer for QST to translate measured statistics into quantum states. In particular, we queried quantum operators to allow for informative representations of quantum data and incorporated the Bures distance into an integrated loss function to better evaluate the state fidelity. Furthermore, we compared the proposed method with two ML-based methods (FCN, CNN) and two traditional methods. Extensive simulations demonstrated the potential of leveraging quantum characteristics into ML-based QST methods. Experiments on IBM quantum devices suggest that ML-based QST exhibits robustness over other methods. However, one should note that our method does not solve the exponential scaling challenge for full tomography of quantum states. In the future, we can incorporate shadow tomography that is able to predict certain observables for quantum states of arbitrary size using a polynomial number of measurement samples~\cite{huang2020predicting}.  We will also focus on investigating the application of the language model to QST with incomplete measurements, quantum detector tomography, and quantum process tomography.
	

	\appendix
	\section{Appendices}

	\begin{table*}[h]
		\centering
		\caption{Comparison of parameters for our method and other methods. Infer-CPU denotes the inference process performed on CPU, while Infer-GPU denotes the inference process performed on GPU.}
		\begin{tabular}{c|cccc|cccc}
			\toprule[1.5pt]
			& \multicolumn{4}{c|}{2 qubit} & \multicolumn{4}{c}{4 qubit}\\
			\midrule[1.5pt]
			Methods& Params &Train-GPU &  Infer-CPU& Infer-GPU& Params & Train-GPU & Infer-CPU & Infer-GPU\\
			\midrule[1.5pt]
			LRE & N/A & N/A & 4.57ms & N/A & N/A & N/A  & 671ms & N/A \\
			\midrule[1pt]
			FCN & 211K & 1.0h & 5.42ms & 1.30ms & 595K  & 1.5h & 20.35ms & 1.88ms \\
			CNN & 1.23M &  3.4h & 8.76ms & 2.37ms & 30.38M & 25.2h & 85.48ms & 10.55ms \\
			QAT  & 144K & 1.9h & 5.84ms & 1.25ms & 152K  & 5.9h & 21.18ms  & 1.96ms\\
			\bottomrule[1.5pt]
		\end{tabular}
		\label{table:infer}
	\end{table*}

	\subsection{An example of QST for 2-qubit states}\label{app:example}
	
	To illustrate the resemblance between data structures resulting from sequences of quantum measurements and linguistic constructs, we present a specific instance involving the estimation of a 2-qubit state through square root measurements. Specifically, we employ measurements derived from Eq.~(19), where each set comprises 4 operators that sum up to the identity matrix, which is also a quantum detector. One may utilize 5 sets of measurements to meet the informational completeness. Consequently, the observed frequencies can be organized into a $5\times4$ matrix $[f_{ij}]$. Here, each element of this matrix, i.e., one measured frequency $f_{ij}$ corresponds to one character, and four frequencies among each detector $\vec{f}_i=[f_{i1},f_{i2},f_{i3},f_{i4}]$ (i.e., the $i$-th row) correspond to a word. Thus, 20 measured frequencies, or five frequency vectors for five detectors, effectively constitute a sentence.  In this regard, measured statistics of structured measurements (i.e., a quantum detector) serve a similar function to a word. Therefore, the process of reconstructing density matrices from measured statistics can be likened to a language translation process.
	
	\subsection{Computational complexity comparison}
	
	In this section, we compare the number of parameters for the proposed method and other conventional methods for comparison. We also compare Training time (GPU) and Inference time (GPU/CPU) to conduct an overall comparison with the existing methods. The simulation platform consists of an 8-core Intel(R) Xeon(R) W-2145 CPU @ 3.70GHz, 64G DDR4 memory size, and Quadro RTX 4000 with 8G. For inference time (CPU/GPU), we conduct 5,000 times reconstructing density matrix and report the average time per inference in Table \ref{table:infer}. 
	
	Our method utilizes similar or fewer parameters compared to the FCN approach. However, the computational overhead of the transformer mechanism may lead to longer training time. Nevertheless, when considering a single inference on CPU, neural network-based approaches show competitive performance over traditional methods, i.e., LRE. The reason is that the inference stage in NN-based methods does not require extensive iterations and back-propagations, and the inference stage mainly involves multiplication and addition operations, whose efficiency can be naturally enhanced via GPU acceleration.

	\subsection{Training strategies and model complexities}\label{subsec:model}
	In ML communities, the final performance of multi-layer NNs for the specific task is greatly influenced by the training strategies and the model complexities~\cite{lecun2015deep}. As NNs-based QST has been intensively explored, it is highly desirable to investigate systematic training policies for the proposed QAT-QST. In particular, numerical simulations using different training strategies and models are implemented for 2-qubit states to provide an instructive tutorial for training the neural network-based QST.
	
	\begin{table}[h]
		\centering
		\caption{Training strategy comparison on the 2-qubit pure states.}\label{tab:train}
		\vspace{0.3cm}
		\renewcommand\arraystretch{1.1}
		\resizebox{0.47\textwidth}{!}{
			\begin{threeparttable}
				\begin{tabular}{cccccccccc}
					\toprule[1.5pt]
					& Opt & BS & LR & Epochs & LRA & WE & WD & Infid\\
					\midrule[1.5pt]
					\multirow{4}{*}{A}&SGD & 256 & 1e-2 & 500 & Step & 0 & No & 1.73e-03\\
					& SGD & 256 & 5e-3 & 500 & Step & 0 & No & 4.91e-03\\
					& SGD & 256 & 1e-3 & 500 & Step & 0 & No & 5.71e-01\\
					& SGD & 256 & 5e-4 & 500 & Step & 0 & No & 7.36e-01\\
					\midrule[1pt]
					\multirow{4}{*}{B}&Adam & 256 & 1e-2 & 500 & Step & 0 & No & 4.64e-05 \\
					& \textbf{Adam} & \textbf{256} & \textbf{5e-3} & \textbf{500} 
					& \textbf{Step} & \textbf{0} & \textbf{No} & \textbf{3.88e-05}\\
					& Adam & 256 & 1e-3 & 500 & Step & 0 & No & 7.55e-05\\
					& Adam & 256 & 5e-4 & 500 & Step & 0 & No & 1.82e-04\\
					\midrule[1pt]
					\multirow{5}{*}{C}&Adam & 64 & 5e-3 & 500 & Step & 0 & No & 6.12e-05\\
					& Adam & 512 & 5e-3 & 500 & Step & 0 & No & 3.16e-05\\
					& Adam & 256 & 5e-3 & 500 & Cosine & 0 & No & 3.25e-05\\
					& \textbf{Adam} & \textbf{256} & \textbf{5e-3} & \textbf{500} 
					& \textbf{Cosine} & \textbf{20} & \textbf{No} & \textbf{3.03e-05}\\
					& Adam & 256 & 5e-3 & 500 & Cosine & 20 & Yes & 3.46e-04\\
					\midrule[1pt]
					\multirow{2}{*}{D}&Adam & 256 & 5e-3 & 1000 & Cosine & 20 & No & 2.66e-05\\
					& Adam & 256 & 5e-3 & 1500 & Cosine & 20 & No & 3.21e-05\\
					\bottomrule[1.5pt]
				\end{tabular}
				\begin{tablenotes}
					\footnotesize
					\item Opt: Optimizer; BS: Batch Size; LR: Learning Rate; LRA: Learning Rate Attenuation; WE: Warm-up Epochs; WD: Weight Decay; Infid: Infidility.
				\end{tablenotes}
			\end{threeparttable}
		}
	\end{table}

	\begin{table}[h]
		\centering
		\caption{Model complexity comparison on the 2-qubit pure states.}\label{tab:mcomplex}
		\vspace{0.3cm}
		\renewcommand\arraystretch{1.1}
		\resizebox{0.47\textwidth}{!}{
			\begin{threeparttable}
				\begin{tabular}{ccccccccc}
					\toprule[1.5pt]
					& $d_L$ & $d_S$ & $d_H$ & $d_{rate}$ & FLOPs & Params & Infid\\
					\midrule[1.5pt]
					A & 2 & 32 & 4 & 8 & 81M & 36.5K & 3.03e-05 \\
					\midrule[1pt]
					\multirow{4}{*}{B}& 4 & 32 & 4 & 8 & 163M & 72.2K & 2.32e-05 \\
					&\textbf{8} & \textbf{32} & \textbf{4} & \textbf{8} 
					& \textbf{326M} & \textbf{143.6K} & \textbf{1.66e-05} \\
					&16 & 32 & 4 & 8 & 651M & 286.4K & 1.13e-05 \\
					&32 & 32 & 4 & 8 & 1.303G & 572.1K & 1.19e-05 \\
					\midrule[1pt]
					\multirow{3}{*}{C}&8 & 64 & 4 & 8 & 1.29G & 565.712K & 1.53e-05 \\
					&8 & 128 & 4 & 8 & 5.16G & 2.246M & 2.13e-05 \\
					&8 & 256 & 4 & 8 & 20.58G & 8.947M & 2.68e-04 \\
					\midrule[1pt]
					\multirow{3}{*}{D}&8 & 32 & 8 & 8 & 326M & 143.6K & 1.56e-05 \\
					&\textbf{8} & \textbf{32} & \textbf{16} & \textbf{8} 
					& \textbf{326M} & \textbf{143.6K} & \textbf{9.81e-06} \\
					&8 & 32 & 32 & 8 & 326M & 143.6K & 2.01e-05 \\
					\midrule[1pt]
					\multirow{2}{*}{E}&8 & 32 & 16 & 4 & 175M & 77.0K & 2.82e-05 \\
					&8 & 32 & 16 & 16 & 628M & 276.7K & 1.48e-05 \\
					\bottomrule[1.5pt]
				\end{tabular}
				\begin{tablenotes}
					\footnotesize
					\item FLOP: floating point operations per second. Params: parameters. 
				\end{tablenotes}
			\end{threeparttable}
		}
	\end{table}
	\textbf{Training Strategies.}
	The training strategies such as optimizer, warm-up strategy, and learning rate attenuation are typical operations that influence the performance of neural network models~\cite{lecun2015deep}. Owing to the cost-effective comparison of different training strategies, we consider the basic QAT-QST model with the number of hidden layers $d_L=2$, the embedding size $d_S=32$, the multi-head size $d_H=4$ and $d_{rate}=8$. In particular, we consider some typical training strategies and implement a bunch of simulations to demonstrate their impact on QST in Table~\ref{tab:train}. 

	In row $A$ and row $B$ of Table~\ref{tab:train}, the value of the learning rate is varied from $0.01$ to $0.0005$ to compare two widely-used optimizers, i.e., SGD with momentum 0.9, and Adam with $\beta1=0.9$ and $\beta2=0.999$. The performance of Adam is superior to that of SGD and reaches its best when the learning rate equals $5\times 10^{-4}$. In row $C$, with increased batch size, the reconstruction accuracy increases at the cost of growing memory consumption. In order to control the memory consumption, we do not apply a batch size of 512 in different scenarios. When applying cosine learning rate attenuation and a warm-up strategy for the first 20 epochs, the model performance increases greatly. In row $D$,  the model performance increases first and then decreases as training epochs ascend. After the exploration, the most suitable training strategies are listed in bold font in Table~\ref{tab:train} row $C$.
	
	Note that most of the training strategies can be expanded on a baseline without changing the model architecture, which means the performance improvement is cost-free in the inference stage. If sufficient training resources are available,  a large batch size and training epochs are preferred to benefit the overall performance of the QAT-QST model. 
	
	\textbf{Model Complexity.} 
	In theory, without considering training strategies, model complexities are positively correlated with the performance of an ML task~\cite{roberts2022principles}. The complexity of QAT-QST is compounded by the input dimension related to the number of qubits, and transformer layers composed of embedding size $d_S$, the multi-head number $d_H$ of attention modules, the expansion rate of FCN $d_{rate}$  and the number of the stacked layers $d_L$. Among these, $d_S$, $d_H$, and $d_{rate}$ control the model's width, while $d_L$ controls the depth of the model. When optimizing hyperparameters, it is essential to comprehensively consider the model size, and computational constraints to achieve a trade-off between performance and efficiency. Thus, we explore combinations of different depths and widths using a grid search strategy~\cite{bergstra2012random}, as shown in Table~\ref{tab:mcomplex}. 
	
	
	We use the basic model in Table~\ref{tab:train} as the baseline, which is trained on the well-designed training settings, presented in row $A$ of Table~\ref{tab:mcomplex}. Then, the variations of model depth and results are displayed in row $B$. With the growth of depth, the model complexities increase linearly, but the performance increases in a reasonable range and decreases at $d_L=32$. Considering the trade-off between the complexity and the performance, we choose $d_L=8$ as an appropriate value. Then, $d_S$ is conducted in iteration with ascending values, and the model performance gets worse in row $C$. Next, we explore the effectiveness of multi-head attention $d_H$ and find that $d_H=16$ realizes the best performance without increased complexity. The results in row $E$ further verify that $d_L=8$, $d_S=32$, $d_H=16$, and $d_{rate}=8$ are the appropriate settings for QAT-QST in this work. Judging from the above results, the performance of our model is most sensitive to the number of layers and least sensitive to the expansion rate of the FCN. This procedure of model exploration is a general and valuable technique, bringing insight into how to design appropriate models for ML-based QST tasks. Additionally, one could incorporate advanced techniques to reduce computation burdens, such as sparse attention~\cite{wang2021spatten} or model pruning~\cite{kim2022learned}, to reduce the computational burden.

	\bibliographystyle{ieeetr}
	\bibliography{bib/mybib}

\end{document}